\documentclass[aps,prl,reprint,superscriptaddress]{revtex4-2}
\usepackage{orcidlink}
\usepackage{amsmath}
\usepackage{siunitx}

\begin{document}

\title{Echo Enhanced Strong Focusing for Coherent Short-Wavelength Radiation}

\author{Jingyuan Zhao}
\affiliation{Department of Engineering Physics, Tsinghua University, Beijing 100084, China}
\author{Xiujie Deng}
\affiliation{Institute for Advanced Study, Tsinghua University, Beijing 100084, China}
\author{Zhilong Pan}
\affiliation{Department of Engineering Physics, Tsinghua University, Beijing 100084, China}
\author{Alexander Wu Chao}
\affiliation{Institute for Advanced Study, Tsinghua University, Beijing 100084, China}
\author{Chuanxiang Tang}
\email{tang.xuh@tsinghua.edu.cn}
\affiliation{Department of Engineering Physics, Tsinghua University, Beijing 100084, China}

\date{\today}

\begin{abstract}
Storage-ring-based fully coherent light sources, 
including steady-state microbunching (SSMB), as well as 
compact seeded FELs driven by laser plasma accelerators, 
typically have relatively large intrinsic energy spreads.
Extending the spectral reach of these facilities toward the 
X-ray regime represents a major challenge,
as existing seeded schemes require rather extreme parameters 
to generate appreciable microbunching at high harmonics.
In this Letter, we propose an echo enhanced strong focusing
scheme that employs transverse-longitudinal coupling together
with the beam echo effect to simultaneously resolve the energy spread 
bottleneck and enable efficient high-harmonic generation. This approach 
substantially relaxes the requirements on both the intrinsic 
energy spread and the transverse emittance, paving the way for 
soft X-ray production using relatively weak laser modulation. 
Based on this scheme, we further present an SSMB storage ring capable 
of generating kW-level average power 6.7 nm soft X-ray radiation.\end{abstract}

\maketitle

High-brightness coherent short-wavelength radiation is an essential tool 
for modern scientific research\cite{Xray}.
Among advanced light sources, free-electron lasers (FELs) provide 
extremely high peak brightness, excellent temporal coherence, 
and ultrashort pulse durations\cite{FELraw, FELF1, FELF2}.
FELs can operate either in the self-amplified spontaneous emission 
(SASE)\cite{SASE} mode, which is capable of reaching extremely short wavelengths, 
or in external seeding modes with improved temporal coherence.
Among various external seeding schemes\cite{HGHG, EEHG, PEHG}, echo-enabled harmonic 
generation (EEHG) has been experimentally demonstrated as an effective 
approach for generating coherent radiation at high harmonics\cite{EEHG75, EEHG26}.

While seeded schemes have achieved remarkable success in linac-based FELs, 
developing coherent light sources in storage rings has become an active 
topic in recent years.
Novel concepts, including steady-state microbunching (SSMB)\cite{SSMBraw, SSMBexp, SSMBexp2}
, storage-ring 
FELs\cite{RingFEL1, ADM} and storage-ring seeded coherent light sources\cite{RingFEL2, RingFEL3}, aim to combine the high 
repetition rate of storage rings with the full coherence of radiation.
Beyond fundamental scientific research, such sources are also highly attractive 
for high-average-power extreme ultraviolet (EUV) lithography applications\cite{EUVlitho}.
Compared with linacs, electron beams in storage rings naturally 
feature low vertical emittances, yet suffer from much larger initial energy spreads.
To generate coherent short-wavelength radiation by scaling 
the EEHG scheme to high harmonics in a ring, the required laser-induced 
energy modulation must be sufficiently large compared with the initial energy spread.
This requires high seed laser power and significantly increases the 
slice energy spread in the radiator.
Furthermore, despite their strong potential for compact light sources, 
FELs based on laser plasma accelerators (LPAs)\cite{LPA0, LPA1, LPAFELseeded} 
face a similar bottleneck due to the 
large initial energy spread under current technological constraints.

To circumvent this limitation, several schemes, including phase-merging enhanced harmonic 
generation (PEHG)\cite{PEHG, PEHG1}, angular dispersion modulation (ADM)\cite{ADM}, and generalized longitudinal 
strong focusing (GLSF)\cite{GLSF, SSMBTLC}, employ transverse-longitudinal coupling during laser modulation.
These schemes exploit the low transverse emittance to
release the requirement on longitudinal slice energy spread.
However, at high harmonics, the exponential suppression of the bunching factor 
inherent to single-stage modulation still requires very low transverse 
emittance and strong transverse-longitudinal coupling.
Such conditions can significantly enhance intra-beam scattering (IBS) effects 
and impose stringent constraints on lattice design and beam control when applied to storage rings.

In this Letter, we propose a novel 4D/6D phase-space manipulation 
scheme exploiting transverse-longitudinal coupling and 
the beam echo effect, enabling robust high-harmonic generation 
even for electron bunches with large energy spread and large transverse emittance.
 The core mechanism proceeds as follows: 
an initial energy modulation is first introduced in a transverse-dispersion section; 
the electron beam then passes another transverse-dispersion section,
 where sharpened longitudinal phase-space striations are formed;
 finally, the beam echo effect is employed to scale the 
microbunching to high harmonics. 
Compared with conventional EEHG, this approach does not require excessively 
strong energy modulation, while also avoiding the stringent 
requirements on ultralow 
transverse emittance and strong transverse-longitudinal 
coupling characteristic of 
PEHG, ADM and GLSF schemes.

\begin{figure}[b]
\includegraphics[width=\columnwidth]{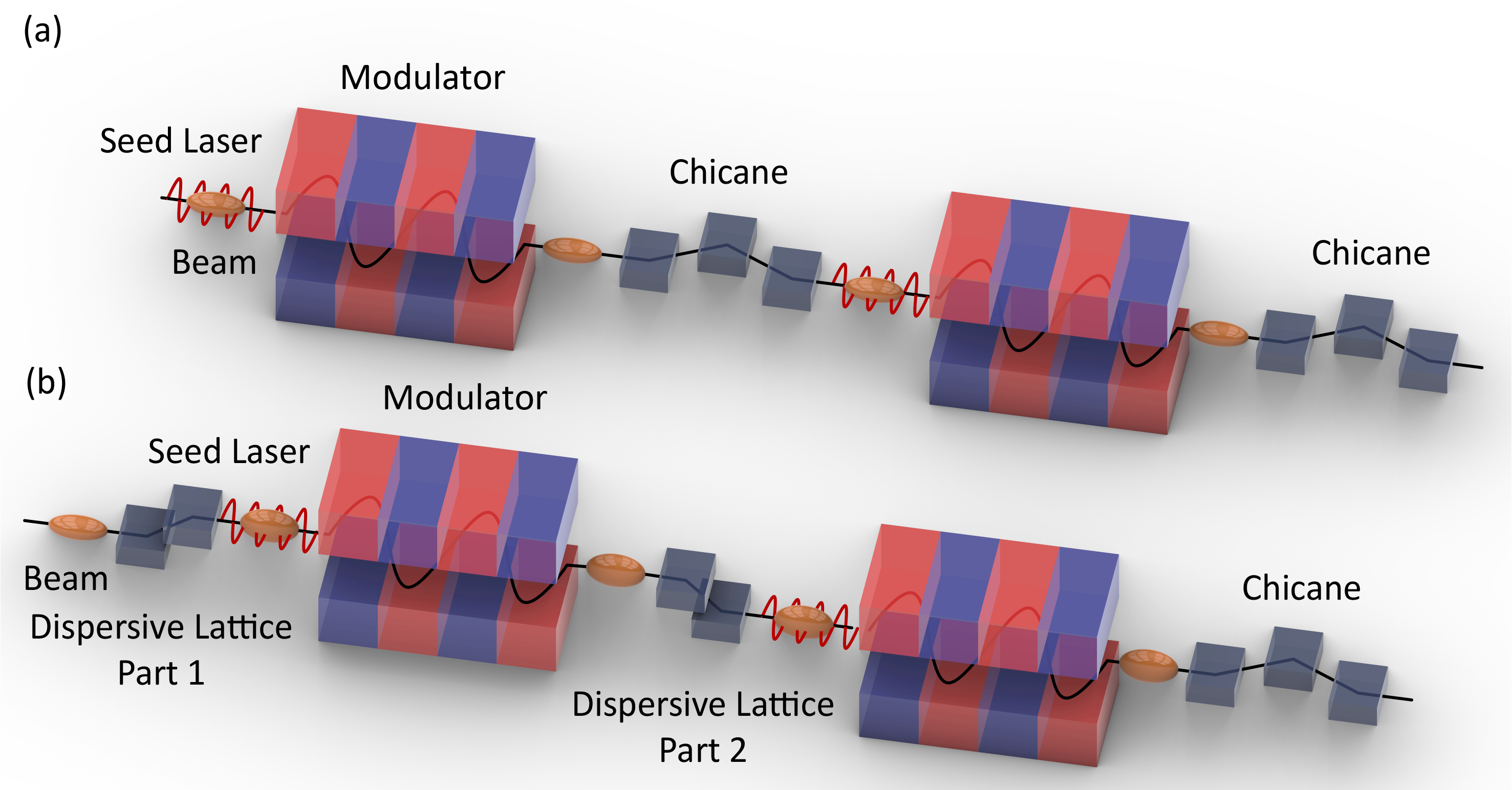}
\caption{\label{fig:setup} (a) Schematic layout of the conventional EEHG scheme, consisting 
of two laser modulators and two chicanes.
(b) Schematic layout of the proposed EESF scheme. Two dispersive lattices are introduced around the first laser modulator to generate vertical-longitudinal coupling.}
\end{figure}

To elucidate the physical mechanism of our proposed scheme, 
we first briefly review the conventional EEHG method. 
The schematic layout of standard EEHG is illustrated in Fig.~\ref{fig:setup}(a). 
The electron beam sequentially passes two stages of laser-induced sinusoidal
energy modulation (with wavenumbers $k_1$ and $k_2$) and chicanes 
(with momentum compactions $R_{56}^{(1)}$ and $R_{56}^{(2)}$). 
Specifically, the first chicane over-compresses the initial 
energy modulation into fine longitudinal phase-space striations.
The second modulation and chicane then convert these striations 
into sharp density microbunching.
The entire 
process relies purely on longitudinal beam dynamics. 
The beam develops density harmonic components at the radiation wavenumber 
$k_r = pk_1 + mk_2$ (where $p$ and $m$ are integers). 
For high-harmonic generation, 
the optimized bunching factor, achieved by properly tuning 
the modulation amplitudes and momentum compaction strengths and setting $p = -1$, 
can be analytically expressed as \cite{EEHGbn}
\begin{equation}\label{EEHGbn}
    b_{-1, m} \approx \frac{0.67}{m^{1/3}}|J_{1}(\xi)|e^{-\frac{\xi^2}{2A_1^2}},
\end{equation}
where $J_1$ is the first-order Bessel function of the first kind, $\xi = A_1\sigma_\delta(k_1R_{56}^{(1)} - k_rR_{56}^{(2)})$, 
$\sigma_\delta$ is the initial rms relative energy spread, and $A_1 = \Delta\delta_1 / \sigma_\delta$ 
is the relative energy modulation amplitude of the first stage, with $\Delta\delta_1$ 
being the relative peak energy modulation amplitude. 
To achieve substantial bunching, $\xi/A_1$ should be kept small while maintaining 
a sufficiently large $J_1(\xi)$, favoring a larger $A_1$. This becomes 
increasingly challenging for beams with large energy spread, where 
the required absolute energy modulation is large, demanding high laser 
power and further increasing the slice energy spread.

To overcome this limitation, we propose utilizing transverse-longitudinal coupling and the beam echo effect
so that the final bunching factor becomes independent of the energy spread. 
The corresponding setup is illustrated in Fig.~\ref{fig:setup}(b), 
which we denote as echo enhanced strong focusing (EESF).
Because the vertical emittance in storage rings is typically much smaller 
than the horizontal one, we consider the vertical-longitudinal phase space $(y,y',z,\delta)$.
Since the laser wavelength is much shorter than the bunch length, 
the beam is approximated as longitudinally uniform.
At the entrance of EESF lattice, the initial phase-space coordinates 
are represented by $(y_0,y_0',z_0,\delta_0)$, with an uncoupled
vertical emittance $\epsilon_y$. 
The beam first passes through a lattice with vertical dispersion which is called part 1. 
Its linear transport matrix is denoted by $t_{ij}$, where the indices $i,j=3\text{--}6$ 
correspond to the phase-space coordinates $(y,y',z,\delta)$.
After this lattice, the vertical Courant-Snyder parameters evolve to $(\alpha_1,\beta_1,\gamma_1)$, 
and the longitudinal coordinate becomes $z_1$.
The transverse-longitudinal coupling strength introduced by part 1 is 
characterized by the vertical dispersion invariant
$\mathcal{H}_1=\gamma_1 t_{36}^2+2\alpha_1 t_{36}t_{46}+\beta_1 t_{46}^2$.
The beam then undergoes a first-stage energy modulation 
$\delta_1=\delta_0+A_1\sigma_\delta \sin(k_1 z_1)$
and then passes a second vertically dispersive lattice, 
referred to as part 2, whose transport matrix elements are denoted by $r_{ij}$ 
with $r_{56}=R_{56}^{(1)}$. 
Finally, the beam passes a second-stage energy modulation with relative amplitude $A_2$, 
followed by a chicane section characterized by $R_{56}^{(2)}$.
The final longitudinal coordinate can 
be expressed as $z_f(y_0, y_0', z_1, \delta_0).$

\begin{figure*}[t]
    \centering
    \includegraphics[width=\textwidth]{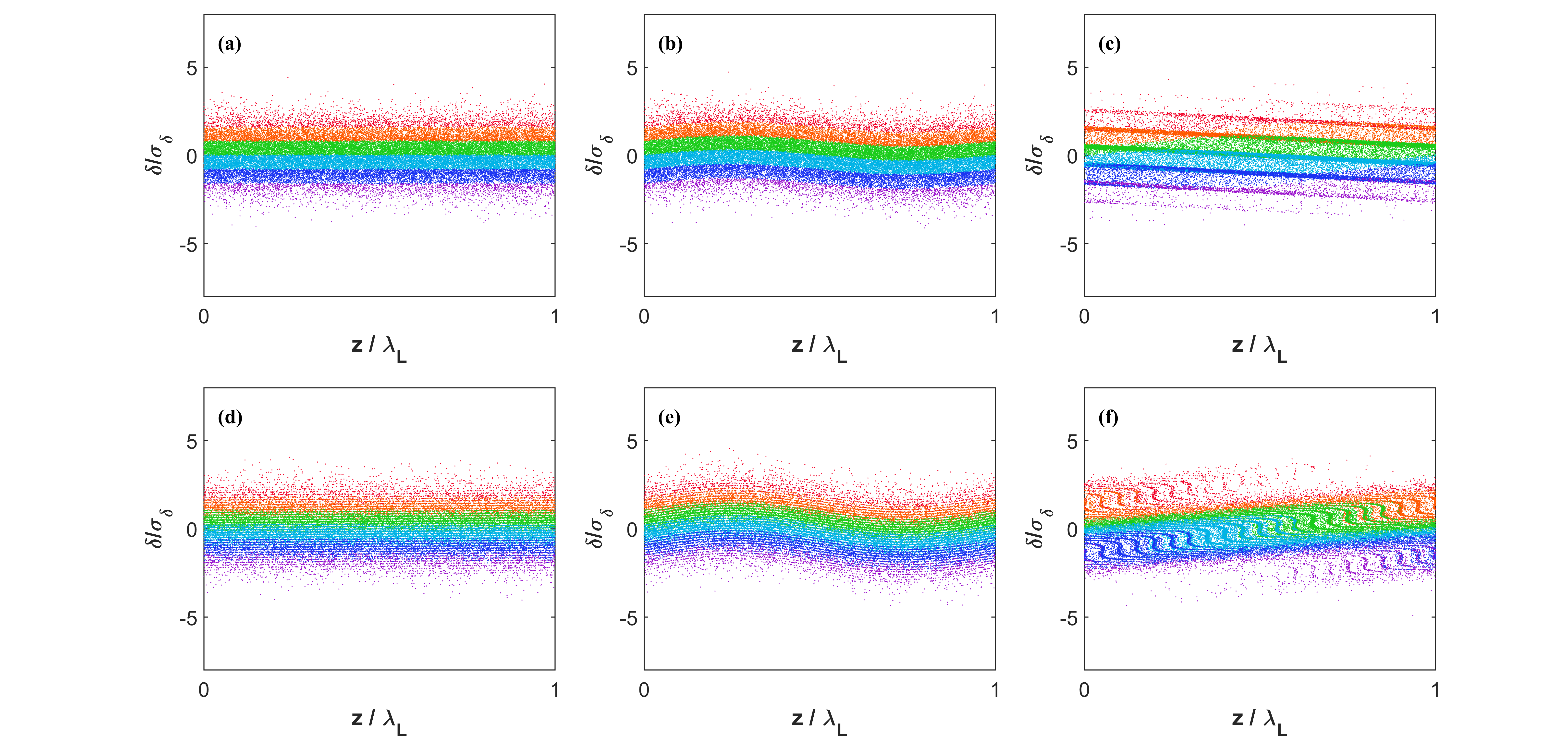}
    \caption{
    Longitudinal phase-space evolution of the beam at different locations in the proposed scheme.
    (a) After the part 1 lattice.
    (b) After the first-stage energy modulation.
    The dynamics in part 2 are separated into two processes:
    (c) removal of the energy-spread dependence of the bunching factor 
    through transverse-longitudinal coupling, and
    (d) first-stage phase-space stretching.
    (e) After the second-stage energy modulation.
    (f) After the second-stage bunch compression.
    The color scale from red to purple represents the vertical position of 
    the electrons before the first-stage modulation, varying from positive to 
    negative values, indicating the correlation between the transverse and 
    longitudinal phase spaces.
    }
    \label{fig:EEGLSFps}
\end{figure*}

\begin{figure}[b]
\includegraphics[width=\columnwidth]{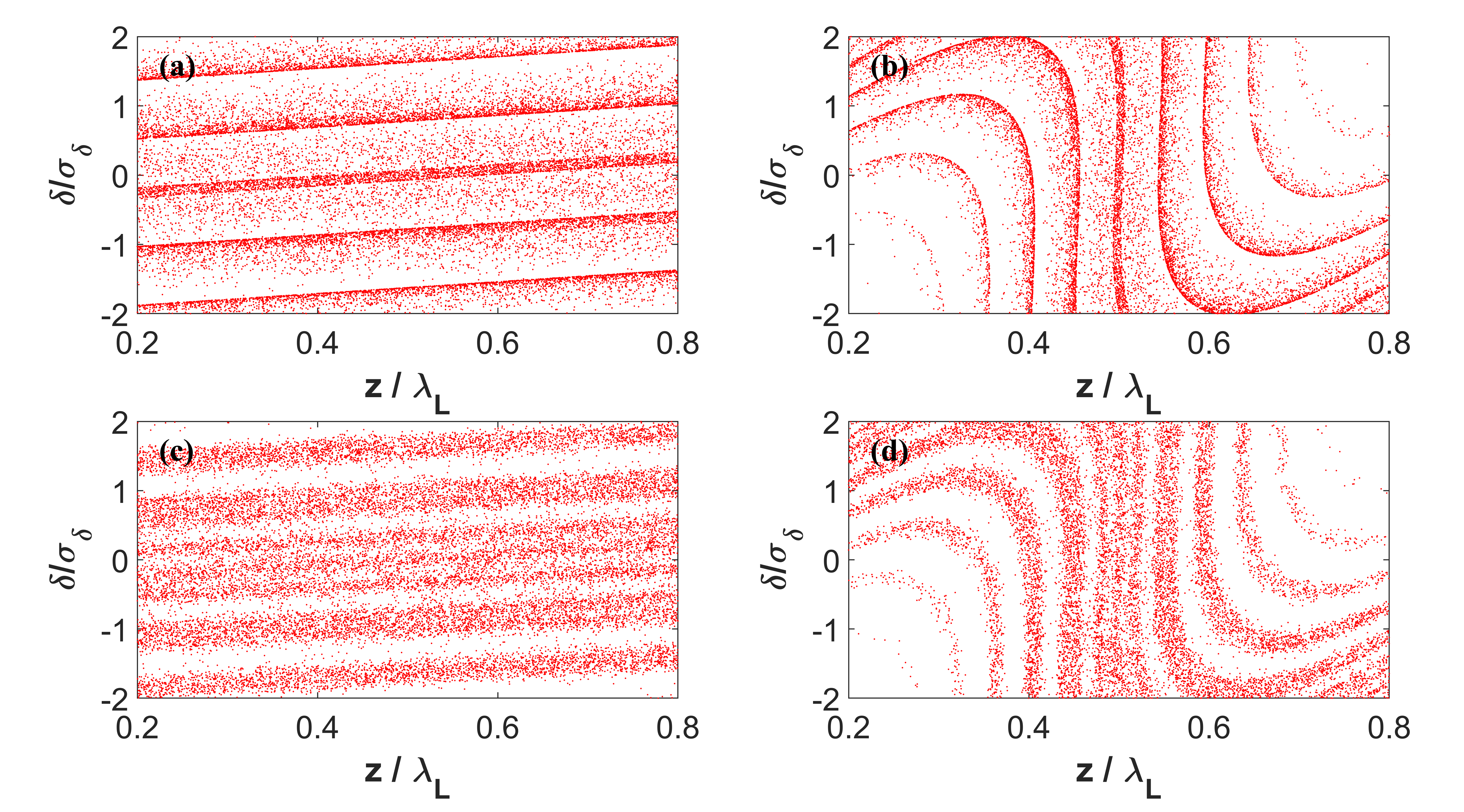}
\caption{\label{fig:EEvsEE} (a),(b) Longitudinal phase spaces of the 
EESF scheme after the first- and second-stage operations, respectively.
(c),(d) Corresponding results for the conventional EEHG scheme.}
\end{figure}

According to Liouville's theorem, the final bunching factor 
at the radiation wavenumber can be written as
\begin{equation}
        b(k_r) = \frac{1}{N_0} |\int f_0(y_0, y_0', \delta_0) e^{-ik_rz_f}dy_0dy_0'dz_1d\delta_0|,
\end{equation}
where $f_0$ is the initial beam distribution function in the vertical-longitudinal 
phase space, assumed to be Gaussian in $(y_0,y_0')$ and $\delta_0$, 
and $N_0$ is the total number of electrons.
Nonvanishing bunching occurs only at the resonant wavenumber
$k_r = p k_1 + m k_2$.
The maximum bunching is obtained for $p=-1$, leading to 
\begin{equation}
   b_{-1, m} = |J_m(Y)J_{1}(\xi)|\exp(-\frac{\epsilon_y}{2}k_1^2\mathcal{W}_y)\exp(-\frac{1}{2}C^2\sigma_\delta^2),
\end{equation}
where $Y = -A_2\sigma_\delta k_rR_{56}^{(2)}$ and $\mathcal{W}_y$ is a lattice-dependent 
coefficient determined by the transport matrices of the two dispersive sections 
and the initial Courant-Snyder parameters. 
The parameter $C$, which plays a central role in eliminating the 
energy-spread-induced suppression of the bunching factor, is given by
\begin{equation}
    C =  k_1(r_{53}t_{36}+r_{54}t_{46}) + k_1R_{56}^{(1)} - k_rR_{56}^{(2)}.
\end{equation}

By properly tuning the dispersive sections in part 2, 
the condition $C=0$ can be achieved, thereby eliminating the energy-spread-dependent 
suppression term. 
The coefficient $\mathcal{W}_y$ can then be minimized, thereby maximizing the bunching factor. 
The minimum attainable value of $\mathcal{W}_y$ is given by
\begin{equation}
    \mathcal{W}_{y, \mathrm{min}} = \frac{\xi^2}{A_1^2k_1^2\sigma_\delta^2\mathcal{H}_1}.
\end{equation}
By optimizing $Y$ to maximize $J_m$, the final bunching factor for $m>4$ becomes
\begin{equation}\label{EESFbn}
b_{-1, m} \approx \frac{0.67}{m^{1/3}}|J_{1}(\xi)|\exp(-\frac{\epsilon_y\xi^2}{2A_1^2\sigma_\delta^2\mathcal{H}_1}),
\end{equation}
which is completely independent of the intrinsic energy spread\cite{SupM}.
This bunching factor exceeds that of conventional EEHG once the condition
$\epsilon_y/\mathcal{H}_1 < \sigma_\delta^2$
is satisfied. 
Such a condition can be readily achieved 
since the transverse beam parameters are generally easier to control than 
the intrinsic energy spread.
Moreover, the exponential suppression term no longer follows the $\exp(-Bn^2\epsilon_y/2A_1^2\sigma_\delta^2\mathcal{H}_1)$ 
scaling characteristic of PEHG-like schemes, where $n$ is the harmonic number and $B \approx 1$. 
As a result, the parameter $\epsilon_y/\mathcal{H}_1$ does not need to decrease 
rapidly with increasing harmonic number, substantially relaxing the 
high-harmonic optimization requirement.
It should be noted that the chicane following the second modulator 
can be replaced by a vertically dispersive lattice without affecting 
the optimal bunching factor given by Eq.~\ref{EESFbn}. Therefore, the 
dispersive sections in parts 1 and 2 alone are sufficient to 
achieve the maximum bunching.

Figure~\ref{fig:EEGLSFps} illustrates the longitudinal phase-space evolution 
of the proposed scheme. 
A relatively weak first-stage energy modulation is used. 
Through the transverse-longitudinal coupling dynamics, 
the sinusoidal modulation in the longitudinal phase space is sharpened.
The sharpened striations is then stretched to high harmonics 
by a large $R_{56}^{(1)}$. 
Finally, the high-harmonic microbunching is revealed through the second-stage 
energy modulation and longitudinal compression.
Fig.~\ref{fig:EEvsEE} compares the detailed longitudinal phase spaces 
of EESF and EEHG before the second-stage modulation and after the full manipulation.
Unlike EEHG, which directly stretches the sinusoidal modulation, EESF 
first sharpens the modulation waveform with the same slope,
 resulting in much clearer final microbunching structures.

The introduction of transverse-longitudinal coupling inevitably leads to 
growth of the vertical emittance. 
Therefore, the coupling strength must be properly controlled to preserve 
the transverse coherence of the radiation. 
At the radiator, the coupled vertical emittance is given by
\begin{equation}
    \epsilon_{y, \mathrm{c}}^2 = \epsilon_y^2 + \epsilon_y\mathcal{H}_1\sigma_\delta^2+ 
    (1 + \frac{A_1^2}{2})\epsilon_y\mathcal{H}_2\sigma_\delta^2 + \frac{A_1^2}{2}\mathcal{H}_1\mathcal{H}_2\sigma_\delta^4,
\end{equation}
where $\mathcal{H}_2=\gamma_2 r_{36}^2+2\alpha_2 r_{36}r_{46}
+\beta_2 r_{46}^2$ characterizes the transverse-longitudinal coupling 
introduced by part 2 with $(\alpha_2,\beta_2,\gamma_2)$ the Courant-Snyder parameters after part 2.
Inspection of the expression for $\mathcal{W}_y$ reveals that it is physically 
equivalent to $\mathcal{H}_2$. 
This expression for the coupled vertical emittance is valid 
only when the bunching factor is maximized.

\begin{figure}[b]
\includegraphics[width=\columnwidth]{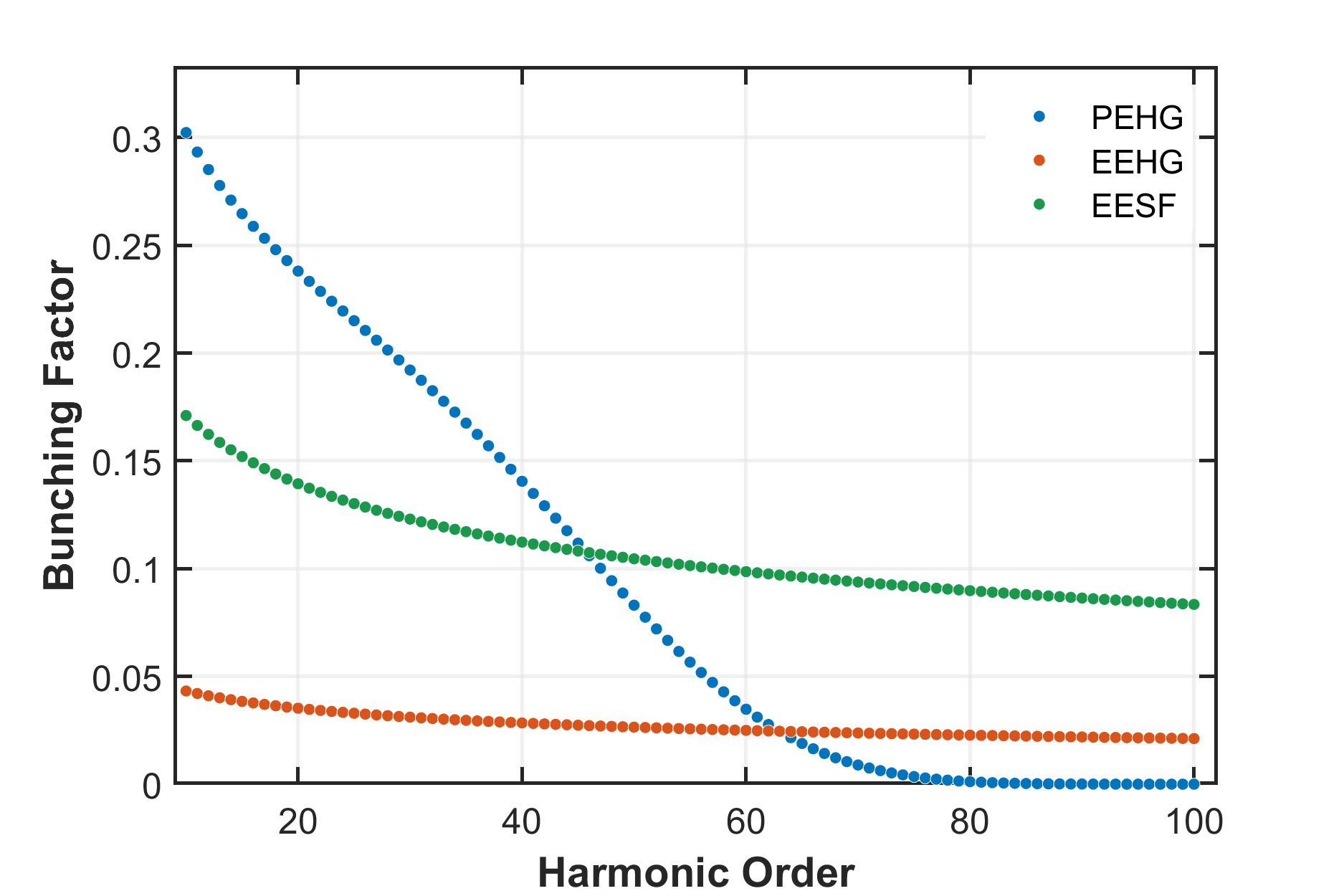}
\caption{\label{fig:bnc} Bunching factor versus harmonic number for PEHG, EEHG and EESF. 
The beam parameters are $\epsilon_y = \SI{5}{pm}$, $\sigma_\delta=1\times10^{-3}$ and $A_1=0.5$. 
For PEHG and EESF, $\mathcal{H}_1$ is limited by the coupled vertical emittance. 
At each harmonic, the lattice parameters of all schemes are optimized to 
maximize the bunching factor.}
\end{figure}

\begin{figure}[b]
\includegraphics[width=\columnwidth]{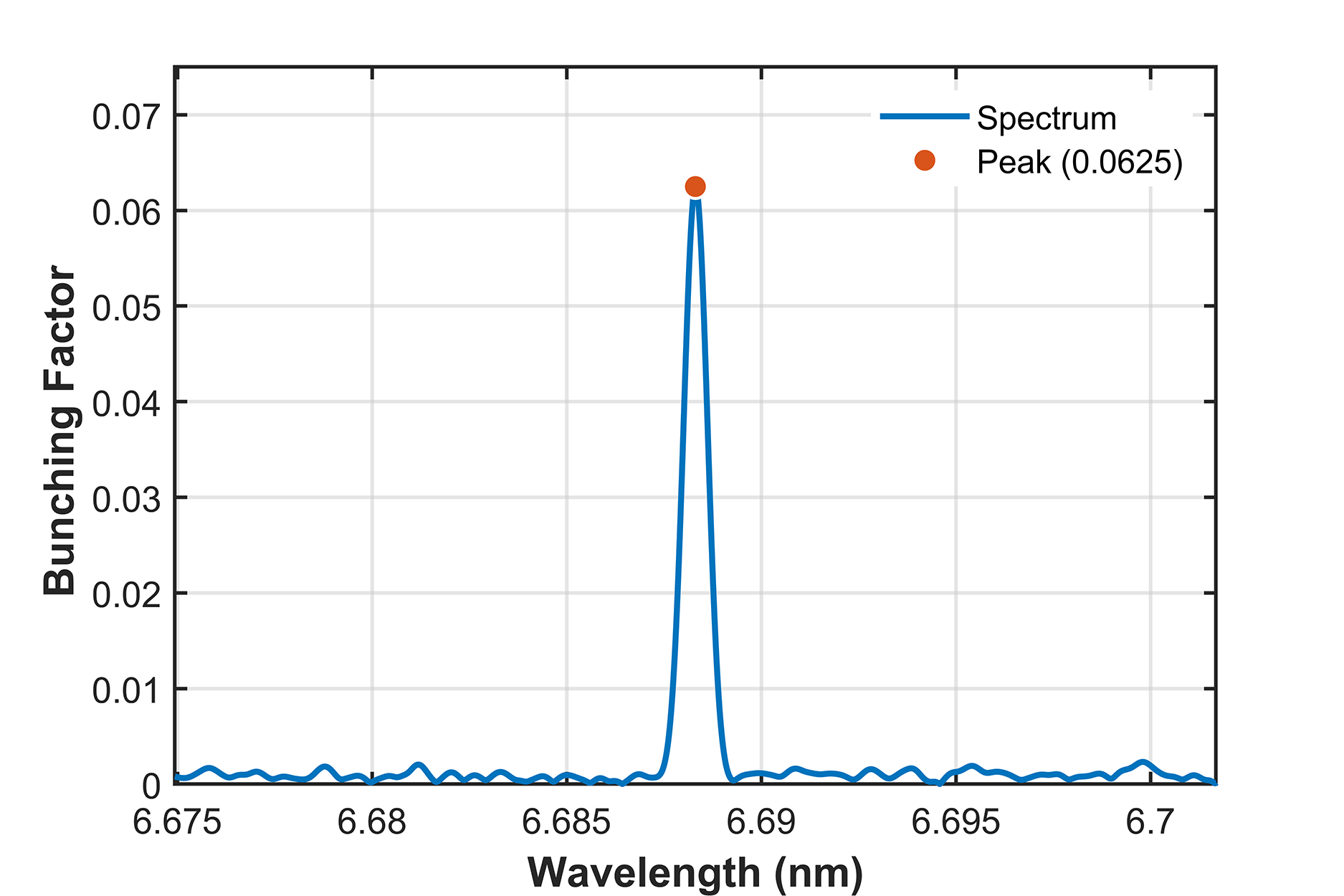}
\caption{\label{fig:SSMBbn} Bunching factor near 6.688 nm at the radiator in the 
EESF SSMB storage ring after 10000-turn tracking.}
\end{figure}

To further investigate the high-harmonic performance under relatively large 
intrinsic energy spread and weak energy modulation, the bunching factors of 
PEHG, EEHG, and EESF are compared. In the calculation, 
the seed laser wavelength is \SI{266}{nm}, $\epsilon_y = \SI{5}{pm}$, 
$\sigma_\delta = 1\times10^{-3}$, and $A_1=0.5$ where $A_1$ corresponds to the first-stage 
energy modulation amplitude in EEHG and EESF and to the energy modulation amplitude in PEHG.
Fig.~\ref{fig:bnc} shows the bunching factor 
for harmonics ranging from the 10th to the 100th.
For PEHG and EESF, 
$\mathcal{H}_1$ is constrained by the requirement that the 
coupled vertical emittance should remain smaller than $\lambda_r/4\pi$,
 where $\lambda_r$ is the radiation wavelength.
Due to the relatively large intrinsic energy spread, $A_1=0.5$ 
already corresponds to a strong energy modulation.
It can be seen that, under such parameters, 
the EESF scheme can generate a bunching factor 
significantly higher than that of EEHG.
Since the vertical emittance is fixed, the bunching factor of the PEHG 
scheme decreases rapidly as the harmonic number increases. 
Since a vertical emittance of \SI{5}{pm} is already small for a 
storage ring, EESF shows better emittance tolerance and stronger 
capability for extending to higher harmonics compared with PEHG.

For the application of the EESF scheme in an SSMB storage ring toward 
high-average-power \SI{6.7}{nm} soft X-ray radiation, we consider a \SI{1.2}{GeV}
storage ring operating with an average beam current of \SI{1}{A} and a peak 
current of \SI{50}{A}. Conventional RF bunches with an rms bunch length of \SI{5}{mm}
are stored in the ring.
The equilibrium 
horizontal emittance $\epsilon_x$ is assumed to be \SI{0.55}{nm}, while 
$\epsilon_y = \SI{30}{pm}$ and $\sigma_\delta = 1\times 10^{-3}$. 
The modulation amplitude parameter is chosen as $A_1=0.2$, 
which can be readily achieved with 
currently available \SI{1030}{nm} optical cavity power\cite{OpticalCavity}.
To ensure transverse coherence, $\mathcal{H}_1$ is 
taken to be \SI{0.01}{m}.
Under these conditions, the calculated bunching factor reaches 0.064.

Based on this parameter set, an EESF insertion lattice together with its inverse 
lattice was designed.
Under these parameters, the vertical IBS must be carefully considered to 
preserve the steady state.
Since transverse-longitudinal coupling is introduced only in 
the EESF insertion section, the IBS growth is dominated by this section\cite{IBSH}.
Calculations show that the resulting IBS enhancement remains moderate 
and can be compensated by additional radiation damping.
Including the sinusoidal energy modulation in the insertion and the 
linear beam dynamics of the storage ring, the entire ring is modeled 
using a transfer matrix formalism. After 10000-turn tracking of a 
test beam in the storage ring, a bunching factor of 0.062 at the radiation 
wavelength is still observed, as shown in Fig.~\ref{fig:SSMBbn}, in good 
agreement with the theoretical prediction.
Under these conditions, the beam can provide an average radiation power of 1000 W.
All parameters adopted here are compatible with state-of-the-art 
accelerator and laser technologies.

In summary, we have proposed a new approach that exploits transverse-longitudinal 
coupling and the beam echo effect to realize microbunching at high harmonics. 
The method exhibits improved robustness 
in the presence of large energy spread, and provides enhanced bunching 
performance compared with existing schemes in 
the high-harmonic regime.
Several collective effects and realistic machine imperfections have not been 
included in the present analysis, including coherent synchrotron radiation (CSR) 
during bunching, lattice imperfections, and fluctuations in the laser modulation system.
These effects should be systematically evaluated 
in the design of realistic experimental implementations.
Notably, under parameters achievable in current fourth-generation storage rings, 
this scheme can produce significant bunching in the hard X-ray regime, 
opening a path toward high-repetition-rate coherent hard X-ray 
radiation from storage rings.
Beyond storage-ring-based fully coherent light sources, 
the proposed scheme may also be applied to existing EEHG FEL facilities, 
where an improvement of the bunching factor can be achieved by modifying 
only the lattice of the bunching section under certain parameter regimes. 
In addition, it may be combined with round-to-flat beam transformations\cite{RTFB} 
in linac-based FEL injectors to further reduce the transverse emittance in one 
dimension and enhance the bunching efficiency.
Moreover, the mechanism may also be relevant to LPA based FELs. In such systems, the 
induced transverse-longitudinal coupling can be naturally exploited in 
combination with transverse gradient undulators (TGUs)\cite{TGU} to improve 
FEL gain.

\vspace{1em}
\noindent\textit{Acknowledgments}---We thank Zizheng Li and Wenxuan Wu 
for valuable discussions on storage ring 
beam dynamics. We also acknowledge discussions with 
Zhuoyuan Liu on FEL dynamics.
This work is supported by the National Key Research 
and Development Program of China (Grant No. 2022YFA1603400),
the National Natural Science Foundation of 
China (Grant No. 12522512) and
the Tsinghua University Initiative Scientific Research Program.

\bibliography{EESF}

\clearpage

\onecolumngrid

\begin{center}
  \textbf{\large Supplemental Material for: Echo Enhanced Strong Focusing for Coherent Short-Wavelength Radiation}
\end{center}

\setcounter{equation}{0}
\setcounter{figure}{0}
\setcounter{table}{0}
\setcounter{page}{1}
\setcounter{section}{0}

\renewcommand{\theequation}{S\arabic{equation}}
\renewcommand{\thefigure}{S\arabic{figure}}
\renewcommand{\thetable}{S\arabic{table}}
\renewcommand{\thesection}{S\arabic{section}}
\section{Detailed Vertical-Longitudinal Phase-Space Manipulation in EESF}

Before the EESF manipulation, we consider an electron bunch characterized by a uniform 
longitudinal current distribution, a transverse emittance $\epsilon_y$, and 
an energy spread $\sigma_\delta$. The initial vertical Courant-Snyder parameters are given by 
$(\alpha_0, \beta_0, \gamma_0)$, and the initial phase-space coordinates are denoted 
by $(y_0, y_0', z_0, \delta_0)$. The initial phase-space distribution can be expressed as
\begin{equation}
    f_0(y_0, y_0', z_0, \delta_0) = \frac{N_0}{(2\pi)^{3/2}\epsilon_y\sigma_\delta}\exp{[-\frac{\gamma_0y_0^2 + 2\alpha_0y_0y_0'+\beta_0y_0'^2}{2\epsilon_y} - \frac{\delta_0^2}{2\sigma_\delta^2}]},
\end{equation}
where $N_0$ is the total electron number.

First, we utilize part 1 to introduce vertical-longitudinal coupling. 
The corresponding transport matrix can be expressed as
\begin{equation}
    M_{\textrm{part 1}} = \begin{bmatrix}
    t_{33} & t_{34} & 0 & t_{36} \\
    t_{43} & t_{44} & 0 & t_{46} \\
    t_{53} & t_{54} & 1 & t_{56} \\
    0 & 0 & 0 & 1 \\
\end{bmatrix}.
\end{equation}
After this manipulation, the phase-space coordinates of the electrons are given by 
$(y_1, y_1', z_1, \delta_0)$, with the corresponding Courant-Snyder parameters being $(\alpha_1, \beta_1, \gamma_1)$.
We then introduce the first-stage energy modulation, which can be expressed as
\begin{equation}
\delta_1 = \delta_0 + A_1\sigma_\delta\sin(k_1z_1),
\end{equation}
where $A_1$ is the relative amplitude of the first-stage energy modulation as defined in the 
main text, and $k_1$ is the wavenumber of the first-stage modulating laser.
Here, $z_1$  can be expressed as
\begin{equation}
    z_1 = z_0 + t_{53}y_0 + t_{54}y_0' + t_{56}\delta_0.
\end{equation}

Subsequently, we introduce part 2, a second vertical dispersion section, to sharpen and 
stretch the striations in the longitudinal phase space.
The transport matrix of part 2 is
\begin{equation}
    M_{\textrm{part 2}} = \begin{bmatrix}
    r_{33} & r_{34} & 0 & r_{36} \\
    r_{43} & r_{44} & 0 & r_{46} \\
    r_{53} & r_{54} & 1 & R_{56}^{(1)} \\
    0 & 0 & 0 & 1 \\
\end{bmatrix},
\end{equation}
where $R_{56}^{(1)}$ is the first-stage momentum compaction.
After this manipulation, the phase-space coordinates of the electrons are given by 
$(y_2, y_2', z_2, \delta_1)$, with the corresponding Courant-Snyder parameters being $(\alpha_2, \beta_2, \gamma_2)$.
Here, $z_2$  can be expressed as
\begin{equation}
    z_2 = z_1 + r_{53}(t_{33}y_0 + t_{34}y_0' + t_{36}\delta_0) + r_{54}(t_{43}y_0 + t_{44}y_0' + t_{46}\delta_0) + R_{56}^{(1)}(\delta_0 + A_1\sigma_\delta\sin(k_1z_1)).
\end{equation}
We then introduce the second-stage energy modulation, which can be expressed as
\begin{equation}
\delta_2 = \delta_1 + A_2\sigma_\delta\sin(k_2z_2 + \phi),
\end{equation}
where $A_2$ is the relative amplitude of the second-stage energy modulation,
 $k_2$ is the wavenumber of the second-stage modulating laser,
  and $\phi$ is an arbitrary phase. 

  Subsequently, the bunch passes through a chicane with a momentum compaction of $R_{56}^{(2)}$, after which 
  the final longitudinal coordinate of the particle, $z_f$, can be expressed as
\begin{equation}
z_f = z_2 + R_{56}^{(2)}(\delta_0 + A_1\sigma_\delta\sin(k_1z_1) + A_2\sigma_\delta\sin(k_2z_2 + \phi)).
\end{equation}
Ultimately, this final coordinate $z_f$ can be expressed in terms of the variables $(y_0, y_0', z_1, \delta_0)$.

\section{Derivation of the Bunching Factor}

We evaluate the bunching factor at the radiation wavenumber $k_r$, which is defined as
\begin{equation}
    b(k_r)  = \frac{1}{N_0} |\int\int\int\int f_f(y_f, y_f', z_f, \delta_f) e^{-ik_rz_f}dy_fdy_f'dz_fd\delta_f|,
\end{equation}
where the subscript $f$ denotes the final state after all manipulation steps.
According to Liouville's theorem, the bunching factor can be converted into an integral over the initial state
\begin{equation}
    b(k_r)  = \frac{1}{N_0} |\int\int\int\int f_0(y_0, y_0', \delta_0) e^{-ik_rz_f}dy_0dy_0'dz_0d\delta_0|.
\end{equation}
Since the bunch is longitudinally uniform, the variable $z_0$ can be replaced by $z_1$.
\begin{equation}
    b(k_r)  = \frac{1}{N_0} |\int\int\int\int f_0(y_0, y_0', \delta_0) e^{-ik_rz_f(y_0, y_0', z_1, \delta_0)}dy_0dy_0'dz_1d\delta_0|.
\end{equation}
For simplicity, we write $z_2$ and $z_f$ as
\begin{equation}
    \begin{split}  
    z_2 &= z_1 + g_0(y_0, y_0') + g_1\delta_0 + R_{56}^{(1)}A_1\sigma_\delta\sin(k_1z_1), \\
    z_f &= z_1 + g_0(y_0, y_0') + g_2\delta_0 + (R_{56}^{(1)} + R_{56}^{(2)})A_1\sigma_\delta\sin(k_1z_1) + R_{56}^{(2)}A_2\sigma_\delta\sin(k_2z_2 + \phi),
    \end{split}
\end{equation}
where $g_0(y_0, y_0') = (r_{53}t_{33} + r_{54}t_{43})y_0 + (r_{53}t_{34} + r_{54}t_{44})y_0'$, 
$g_1 = r_{53}t_{36} + r_{54}t_{46} + R_{56}^{(1)}$ and $g_2 = g_1 + R_{56}^{(2)}$.
Then, $e^{-ik_rz_f}$ can be written as
\begin{equation}
    e^{-ik_rz_f} = b_1b_2b_3 = e^{-ik_r(z_1 + g_0 + g_2\delta_0)}e^{-ik_r(R_{56}^{(1)} + R_{56}^{(2)})A_1\sigma_\delta\sin(k_1z_1)}e^{-ik_rR_{56}^{(2)}A_2\sigma_\delta\sin(k_2z_2 + \phi)}, 
\end{equation}
where $b_1$, $b_2$, and $b_3$ represent the three exponential terms in the expression, respectively.
Considering $b_3$, it can be decomposed as
\begin{equation}
    \begin{split}
        b_3 &= e^{-ik_rR_{56}^{(2)}A_2\sigma_\delta\sin(k_2z_2 + \phi)}\\
            &= \sum_{m = -\infty}^{\infty}J_m(-k_rR_{56}^{(2)}A_2\sigma_\delta)e^{im(k_2z_2 + \phi)}\\
            &= \sum_{m = -\infty}^{\infty}J_m(-k_rR_{56}^{(2)}A_2\sigma_\delta)e^{im(k_2(z_1 + g_0 + g_1\delta_0) + \phi)}e^{imk_2R_{56}^{(1)}A_1\sigma_\delta\sin(k_1z_1)},
    \end{split}
\end{equation}
where $J_m$ is the $m$-th order Bessel function of the first kind.
Then, we consider $b_2b_3$, 
\begin{equation}
    \begin{split}
        b_2b_3 &= e^{-ik_r(R_{56}^{(1)} + R_{56}^{(2)})A_1\sigma_\delta\sin(k_1z_1)}e^{-ik_rR_{56}^{(2)}A_2\sigma_\delta\sin(k_2z_2 + \phi)}\\
               &= \sum_{m = -\infty}^{\infty}J_m(-k_rR_{56}^{(2)}A_2\sigma_\delta)e^{im(k_2(z_1 + g_0 + g_1\delta_0) + \phi)}\sum_{p = -\infty}^{\infty}J_p(mk_2R_{56}^{(1)}A_1\sigma_\delta - k_r(R_{56}^{(1)} + R_{56}^{(2)})A_1\sigma_\delta)e^{ipk_1z_1}\\
               &= \sum_{m = -\infty}^{\infty}\sum_{p = -\infty}^{\infty}J_m(Y_{A_2})J_p(Y_{A_1})e^{imk_2g_0}e^{imk_2g_1\delta_0}e^{im\phi}e^{i(mk_2+pk_1)z_1},
    \end{split}    
\end{equation}
where $Y_{A_2} = -k_rR_{56}^{(2)}A_2\sigma_\delta$, $Y_{A_1} = (mk_2R_{56}^{(1)} - k_r(R_{56}^{(1)} + R_{56}^{(2)}))A_1\sigma_\delta$.
Finally, $e^{-ik_rz_f}$ can be written as 
\begin{equation}
    e^{-ik_rz_f} = \sum_{m = -\infty}^{\infty}\sum_{p = -\infty}^{\infty}J_m(Y_{A_2})J_p(Y_{A_1})e^{i(mk_2-k_r)g_0}e^{i(mk_2g_1-k_rg_2)\delta_0}e^{im\phi}e^{i(mk_2+pk_1-k_r)z_1}.
\end{equation}
Considering the integration with respect to $z_1$, the term $e^{i m \phi}$ does not affect the result, and the integral is non-zero only when the following condition is satisfied
\begin{equation}
    k_r = mk_2 + pk_1.
\end{equation}
The bunching factor is maximized at $p = \pm 1$. 
To ensure that $R_{56}^{(1)}$ and $R_{56}^{(2)}$ have the same sign, we choose $m > 0$ and $p = -1$.
Consequently, the bunching factor can be expressed as
\begin{equation}
        \begin{split}
            b_{-1, m} &= |J_m(Y_{A_2})J_{-1}(Y_{A_1})|\int\int\frac{1}{2\pi\epsilon_y}e^{-\frac{\gamma_0y_0^2 + 2\alpha_0y_0y_0'+\beta_0y_0'^2}{2\epsilon_y}}e^{ik_1g_0}dy_0dy_0'\int \frac{1}{\sqrt{2\pi}\sigma_\delta}e^{- \frac{\delta_0^2}{2\sigma_\delta^2}}e^{iC\delta_0}\\
                     &= |J_m(Y_{A_2})J_{-1}(Y_{A_1})|e^{-\frac{1}{2}\epsilon_yk_1^2\mathcal{W}_y}e^{-\frac{1}{2}C^2\sigma_\delta^2},
        \end{split} 
\end{equation}
where $Y_{A_1}$ is consistent with the definition of $\xi = A_1\sigma_\delta(k_1R_{56}^{(1)} - k_rR_{56}^{(2)})$ in the main text.
The two lattice-dependent constant terms $\mathcal{W}_y$ and $C$ are given by
\begin{equation}
    \begin{split}
        \mathcal{W}_y &= \beta_0 (r_{53}t_{33} + r_{54}t_{43})^2 - 2\alpha_0(r_{53}t_{33} + r_{54}t_{43})(r_{53}t_{34} + r_{54}t_{44}) + \gamma_0(r_{53}t_{14} + r_{54}t_{44})^2, \\
        C &= k_1(r_{53}t_{36}+r_{54}t_{46}) + k_1R_{56}^{(1)} - k_rR_{56}^{(2)}.
    \end{split} 
\end{equation}
For high-order harmonics $m > 4$, by setting $A_2\sigma_\delta k_rR_{56}^{(2)} = m + 0.81m^{1/3}$, the bunching factor reaches its maximum value, 
which can be expressed as 
\begin{equation}
    b_{-1, m} \approx \frac{0.67}{m^{1/3}}|J_{1}(\xi)|e^{-\frac{1}{2}\epsilon_yk_1^2\mathcal{W}_y}e^{-\frac{1}{2}C^2\sigma_\delta^2}.
\end{equation}
By adjusting $M_{\textrm{part 2}}$ to satisfy $C = 0$, the bunching factor becomes independent of the beam energy spread
\begin{equation}
    b_{-1, m} \approx \frac{0.67}{m^{1/3}}|J_{1}(\xi)|e^{-\frac{1}{2}\epsilon_yk_1^2\mathcal{W}_y}.
\end{equation}

\section{Optimization of the Bunching Factor}

Maximizing the bunching factor is equivalent to minimizing the $\mathcal{W}_y$ function under the conditions that $C = 0$ and $M_{\textrm{part 1}}$ is known.
Here, we define the following vectors and matrixs
\begin{equation}
    \begin{split}
        \mathbf{R} &= \begin{bmatrix}
            r_{53}\\
            r_{54}
        \end{bmatrix},\\
        \mathbf{D}_1 &= \begin{bmatrix}
            t_{36}\\
            t_{46}
        \end{bmatrix},\\
        \mathbf{M}_1 &= \begin{bmatrix}
            t_{33}& t_{34}\\
            t_{43}& t_{44}
        \end{bmatrix},\\
        \mathbf{T}_0 &= \begin{bmatrix}
            \beta_0& -\alpha_0\\
            -\alpha_0& \gamma_0
        \end{bmatrix},\\
        \mathbf{T}_1 &= \begin{bmatrix}
            \beta_1& -\alpha_1\\
            -\alpha_1& \gamma_1
        \end{bmatrix}.
    \end{split}
\end{equation}
At this point, $\mathcal{W}_y$ can be expressed as
\begin{equation}
    \mathcal{W}_y = (\mathbf{R}^T\mathbf{M}_1)\mathbf{T}_0(\mathbf{M}_1^T\mathbf{R}) = \mathbf{R}^T(\mathbf{M}_1\mathbf{T}_0\mathbf{M}_1^T)\mathbf{R} = \mathbf{R}^T\mathbf{T}_1\mathbf{R},
\end{equation}
where the superscript $T$ represents the matrix transpose.
$C = 0$ can be expressed as
\begin{equation}
    \mathbf{R}^T\mathbf{D}_1 = -\frac{\xi}{A_1\sigma_\delta k_1}.
\end{equation}
Since the matrix $\mathbf{T}_1$ is positive definite, $\mathcal{W}_y$ can reach its minimum under the constraint $C = 0$, which is given by
\begin{equation}
    \mathcal{W}_y(min) = \frac{\xi^2}{A_1^2\sigma_\delta^2k_1^2(\mathbf{D}_1^T\mathbf{T}_1^{-1}\mathbf{D}_1)}.
\end{equation}
At this point, the vector $\mathbf{R}$ corresponding to the minimization of $\mathcal{W}_y$ is given by
\begin{equation}\label{eq1}
    \mathbf{R} = -\frac{\xi}{A_1\sigma_\delta k_1} \frac{\mathbf{T}_1^{-1}\mathbf{D}_1}{ \mathbf{D}_1^T\mathbf{T}_1^{-1}\mathbf{D}_1}.
\end{equation}
Here, $\mathbf{D}_1^T\mathbf{T}_1^{-1}\mathbf{D}_1$ is the vertical dispersion invariant $\mathcal{H}_1=\gamma_1 t_{36}^2+2\alpha_1 t_{36}t_{46}+\beta_1 t_{46}^2$
introduced by part 1.
Therefore, the maximized bunching factor is given by
\begin{equation}
    b_{-1, m} \approx \frac{0.67}{m^{1/3}}|J_{1}(\xi)|\exp(-\frac{\epsilon_y\xi^2}{2A_1^2\sigma_\delta^2\mathcal{H}_1}).
\end{equation}

\section{Transverse Emittance Growth Induced by Transverse-Longitudinal Coupling}

Here, we first consider the physical meaning of the $\mathcal{W}_y$ function.
We define the following matrices and vectors
\begin{equation}
    \begin{split}
    \mathbf{D}_2 &= \begin{bmatrix}
        r_{36} \\
        r_{46}
    \end{bmatrix},\\
    \mathbf{M}_2 &= \begin{bmatrix}
        r_{33} & r_{34} \\
        r_{43} & r_{44}
    \end{bmatrix},\\
    \mathbf{T}_2 &= \begin{bmatrix}
            \beta_2& -\alpha_2\\
            -\alpha_2& \gamma_2
        \end{bmatrix},\\
                \mathbf{S} &= \begin{bmatrix}
            0& 1\\
            -1& 0
        \end{bmatrix}.
    \end{split}
\end{equation}
From the symplectic condition of the transport matrix, we have
\begin{equation}\label{eq2}
    \mathbf{R} = \mathbf{M}_2^T\mathbf{S}\mathbf{D}_2.
\end{equation}
Then, $\mathcal{W}_y$ can be written as 
\begin{equation}
    \begin{split}
    \mathcal{W}_y &= \mathbf{R}^T\mathbf{T}_1\mathbf{R}\\
                  &= \mathbf{D}_2^T\mathbf{S}^T\mathbf{M}_2\mathbf{T}_1\mathbf{M}_2^T\mathbf{S}\mathbf{D}_2\\
                 &= \mathbf{D}_2^T\mathbf{T}_2\mathbf{D}_2\\
                 &= \mathcal{H}_2,
    \end{split}
\end{equation}
where $\mathcal{H}_2=\gamma_2 r_{36}^2+2\alpha_2 r_{36}r_{46}
+\beta_2 r_{46}^2$ characterizes the transverse-longitudinal coupling 
introduced by part 2.

We consider the coupled vertical emittance of the bunch at the end of part 2.
Since no further coupling is introduced downstream, the projected emittance 
here is identical to that at the location where the bunching factor is generated. 
Expressing the vertical coordinates of the electrons in vector form, we have
\begin{equation}
    \mathbf{Y} = \begin{bmatrix}
        y_2 \\
        y_2'
    \end{bmatrix} = 
    \mathbf{Y}_\beta + (\mathbf{M}_2\mathbf{D}_1 + \mathbf{D}_2)\delta_0 + \mathbf{D}_2A_1\sigma_\delta\sin(k_1z_1),
\end{equation}
where $\mathbf{Y}_\beta$ is the electron's intrinsic vertical betatron oscillation coordinate.
The coupled vertical emittance can be written as 
\begin{equation}
    \epsilon_{y, c}^2 = \det \langle \mathbf{Y}\mathbf{Y}^T \rangle,
\end{equation}
where $\langle \mathbf{Y}\mathbf{Y}^T \rangle$ can be written as 
\begin{equation}
    \langle \mathbf{Y}\mathbf{Y}^T \rangle = \epsilon_y \mathbf{T}_2 + \sigma_\delta^2(\mathbf{M}_2\mathbf{D}_1 + \mathbf{D}_2)(\mathbf{M}_2\mathbf{D}_1 + \mathbf{D}_2)^T + \frac{A_1^2\sigma_\delta^2}{2}\mathbf{D}_2\mathbf{D}_2^T.
\end{equation}
Then $\det \langle \mathbf{Y}\mathbf{Y}^T \rangle$ can be written as 
\begin{equation}
    \det \langle \mathbf{Y}\mathbf{Y}^T \rangle = \epsilon_y^2 + \epsilon_y\sigma_\delta^2\mathcal{H}_{\mathrm{tot}}
 + \frac{A_1^2\sigma_\delta^2}{2}\mathcal{H}_2 + \frac{A_1^2\sigma_\delta^4}{2}(\mathcal{H}_{\mathrm{tot}}\mathcal{H}_2 - ((\mathbf{M}_2\mathbf{D}_1 + \mathbf{D}_2)^T\mathbf{T}_2^{-1}\mathbf{D}_2)^2),
\end{equation}
where $\mathcal{H}_{\mathrm{tot}} = (\mathbf{M}_2\mathbf{D}_1 + \mathbf{D}_2)^T\mathbf{T}_2^{-1}(\mathbf{M}_2\mathbf{D}_1 + \mathbf{D}_2) = \mathcal{H}_1 + \mathcal{H}_2 + 2\mathbf{D}_2^T\mathbf{T}_2^{-1}\mathbf{M}_2\mathbf{D}_1$.
Based on the symplecticity of the transport matrix, it can be proven that 
\begin{equation}
\mathcal{H}_{\mathrm{tot}}\mathcal{H}_2 - ((\mathbf{M}_2\mathbf{D}_1 + \mathbf{D}_2)^T\mathbf{T}_2^{-1}\mathbf{D}_2)^2 = (\mathbf{R}^T\mathbf{D}_1)^2.
\end{equation}
Under the condition of maximizing the bunching factor, we have $\mathbf{D}_2^T\mathbf{T}_2^{-1}\mathbf{M}_2\mathbf{D}_1 = 0$ from Eqs.~\eqref{eq1} and \eqref{eq2}
and $(\mathbf{R}^T\mathbf{D}_1)^2 = \mathcal{H}_1\mathcal{H}_2$.
At this point, the coupled vertical emittance can be written as
\begin{equation}
    \epsilon_{y, \mathrm{c}}^2 = \epsilon_y^2 + \epsilon_y\mathcal{H}_1\sigma_\delta^2+ 
    (1 + \frac{A_1^2}{2})\epsilon_y\mathcal{H}_2\sigma_\delta^2 + \frac{A_1^2}{2}\mathcal{H}_1\mathcal{H}_2\sigma_\delta^4.
\end{equation}

\end{document}